\RequirePackage{fix-cm}
\documentclass[smallextended]{svjour3}       
\smartqed  
\usepackage{amssymb}
\usepackage{graphicx}

\usepackage{float}
\usepackage{flushend}
\usepackage{lipsum}
\usepackage[english]{babel}
\usepackage{graphicx}
\usepackage{cite}

\newcommand{\BBB}{\vspace*{-\bigskipamount}}
\usepackage{framed}
\usepackage{balance}
\usepackage{mathtools}
\usepackage{seqsplit}
\usepackage{amsmath}
\usepackage{algorithmic}
\usepackage{array}
\usepackage{wrapfig}
\usepackage{graphicx}
\usepackage{times}
\usepackage{setspace}
\usepackage[caption=false]{subfig}
\usepackage{tabularx,tikz}
\usepackage{tikz-cd}
\usetikzlibrary{matrix,shapes,arrows,positioning,chains, calc}
\usepackage[linesnumbered,ruled,vlined]{algorithm2e}
\usetikzlibrary{shapes.arrows,chains}
\usetikzlibrary{backgrounds}
\tikzstyle{mybox} = [draw=white,   rectangle]
\tikzstyle{fancytitle} =[fill=red, text=white]
\newcommand\newcaptionstyle[2]{%
  \expandafter\ifx\csname caption@@#1\endcsname\relax
    \defcaptionstyle{#1}{#2}%
  \else
    \PackageError{caption}{Caption style `#1' already defined}{}%
  \fi}
\newcommand\defcaptionstyle[2]{%
  \@namedef{caption@@#1}{#2}}

\begin{document}

\title{Vehicle Authentication via Monolithically Certified Public Key and Attributes
\thanks{\textbf{Accepted in Wireless Networks, June 2015.} \\ Partially supported by Rita Altura Trust Chair in Computer Sciences,
Lynne and William Frankel Center for Computer Sciences, Israel Science Foundation (grant number 428/11). Partially supported by fundings from Polish National Science Center (decision number DEC-2013/09/B/ST6/02251). The work of Michael Segal has been supported by General Motors Corporation.}}


\author{Shlomi Dolev         \and
        \L ukasz Krzywiecki  \and
        Nisha Panwar         \and
        Michael Segal  
}


\institute{Shlomi Dolev \at
           Department of Computer Science, Ben-Gurion University of the Negev, Israel. \\
              Tel.: +972-8-6472718\\
              Fax: +972-8-6477650\\
              \email{dolev@cs.bgu.ac.il}           
           \and
           \L ukasz Krzywiecki \at
           Institute of Mathematics and Computer Science, Wroclaw University of Technology, Poland.\\
           Tel.: +48-71-320-3048 \\
           \email {lukasz.krzywiecki@pwr.wroc.pl}
           \and
           Nisha Panwar \at
           Department of Computer Science, Ben-Gurion University of the Negev, Israel. \\
              Tel.: +972-54-6542537\\
              \email{panwar@cs.bgu.ac.il}
           \and
           Michael Segal \at
           Department of Communication Systems Engineering, Ben-Gurion University of the Negev, Israel.\\
           Tel.: +972-8-6477234\\
              Fax: +972-8-6472883\\
           \email {segal@cse.bgu.ac.il}
}

\date{Received: date / Accepted: date}

\maketitle
\begin{abstract}
Vehicular networks are used to coordinate actions among vehicles in traffic by the use of wireless transceivers (pairs of transmitters and receivers). Unfortunately, the wireless communication among vehicles is vulnerable to security threats that may lead to very serious safety hazards. In this work, we propose a viable solution for coping with Man-in-the-Middle attacks. Conventionally, Public Key Infrastructure (PKI) is utilized for a secure communication with the pre-certified public key. However, a secure vehicle-to-vehicle communication requires additional means of verification in order to avoid impersonation attacks. To the best of our knowledge, this is the first work that proposes to certify both the public key and out-of-band sense-able static attributes to enable mutual authentication of the communicating vehicles. Vehicle owners are bound to preprocess (periodically) a certificate for both a public key and a list of fixed unchangeable attributes of the vehicle. Furthermore, the proposed approach is shown to be adaptable with regards to the existing authentication protocols. We illustrate the security verification of the proposed protocol using a detailed proof in Spi calculus.
\end{abstract}

\section{Introduction}
\label{introduction}
Security is a major concern in a connected vehicular network. On one hand, the wireless, ad-hoc, and mobile communication imply security threats, while on the other hand, require perfectly reliable communication, as errors have immediate hazardous implications~\cite{dblp22}. The Intelligent Transportation Systems (ITS) has been regulated as per the standard IEEE 1609 Dedicated Short Range Communication (DSRC)~\cite{dict} and IEEE 802.11p Wireless Access for Vehicular Environment (WAVE)~\cite{zerop}. Also, the security configurations have been standardized as IEEE 1609.2~\cite{vaved} for the online security and ISO 26262~\cite{citeulike} for the functional risk assessment during automotive life cycle. While vehicles move in a predictable road topology, maneuvering among the vehicles is somewhat unpredictable. For example, the vehicle ordering is changed dynamically along the road and over time. Identifying a vehicle is crucially important in the scope of establishing secure communication with passing by vehicles. In particular, using public key infrastructure to establish secure sessions among the moving vehicles is not secure against Man-in-the-Middle (MitM) attacks. Therefore, we propose to modify the conventional certificate structure and facilitate vehicle to vehicle authentication through certificate exchange.

%

\medskip\noindent\textbf{Applications of vehicular networks.}
Gaining on-road safety and efficient traffic management are two prime goals in the use of vehicular networks which is gradually penetrating into the Internet of Things (IoT) communication paradigm. A survey on trust management for IOT~\cite{toit} and the available candidate Internet Engineering Task Force (IETF) solutions across the network layers has been given in~\cite{iettff,persiot}. Smart vehicles may exchange information concerning road scenario with each other to help manage the traffic and to address safety concerns~\cite{DBLP:journals/cn/GerlaK11}. For example, data sharing, remote resource access, payments on the go, a notification on the occurrence of an accident or a traffic jam ahead may assist the approaching vehicles to optimize their time and energy resources. In the very near future, vehicles will interact with several other vehicles to coordinate actions~\cite{ten} and to provide heterogenous media services~\cite{peer2p}. Recently, there have been a great deal of interest to integrate cloud services with the dynamic vehicular communication. Considering the scalability issues and rapid data exchange in vehicular networks in~\cite{cybphy} authors have shown an integrated secure mobile cloud computing~\cite{wei,ali,gaali,securecloud} on top of dynamically scattered cyber-physical vehicle networks. Evidently, these kind of ubiquitous services and real-time applications would definitely require an extended spectrum capacity along with the high-speed gigabit data transfer. Therefore, a great deal of research is being directed to materialize a new cellular networking paradigm termed as 5G~\cite{hetro5g,dtwod}, which would essentially realize the vision of next-generation ad-hoc networks.

%

Several major projects~\cite{maj}, for example, Car2Car-Communication Consortium~\cite{cart}, Cartalk~\cite{tak}, Network on Wheels~\cite{whel}, Vehicle Infrastructure Integration~\cite{vint}, Partners for Advanced Transportation Technology~\cite{calp}, Secure Vehicular Communication~\cite{sevo}, E-safety Vehicle Intrusion Protected Applications~\cite{evi} were conducted in order to initiate, develop and standardize the vehicle network operation. These projects were funded by national governments and accomplished by a joint venture of automobile companies, universities and research organizations. Currently, the vehicle communication research is rapidly trending towards the security aspects~\cite{stand,gpspoof}. Therefore, the focus in this paper is to provide secure wireless communication that is secure against any impersonation attacks by a third party.

\begin{figure}[t!]
  \begin{center}
    \includegraphics[width=10cm]{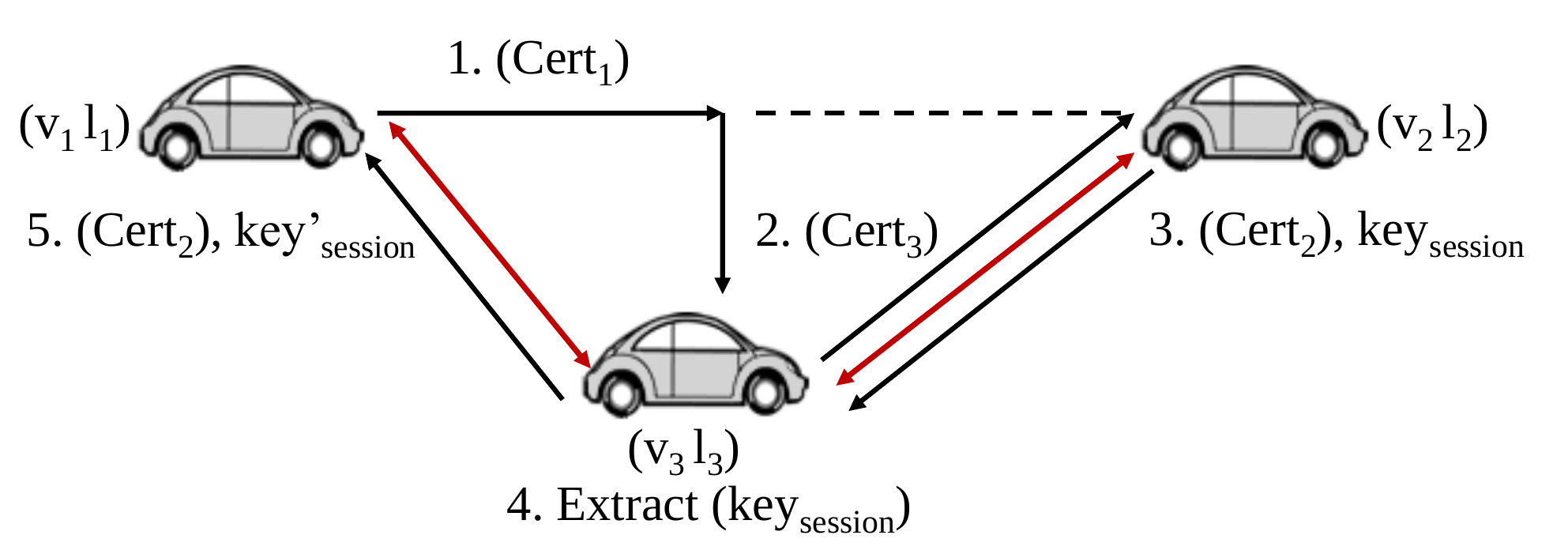}
  \end{center}
  \caption{Attack scenario with the existing PKI.}
  \label{figure:psf}
\end{figure}

\medskip\noindent\textbf{Problem statement.} Public key infrastructure has a severe disadvantage when coping with MitM attacks not only in the scope of vehicle networks. In the common practice public keys are signed by the authorities and can be verified by the receiver. In the scope of vehicle (ad-hoc) networks, secure interaction among the peer vehicles should be established rapidly without any third party assistance. Thus, no interaction with the Certificate Authority (CA) during the session key exchange is feasible and an impersonation attack among the moving vehicles is possible. The following scenario demonstrates a typical MitM attack as shown in Figure~\ref{figure:psf}.

The scenario starts when a vehicle $v_1$ tries to securely communicate with $v_2$ and requests for the public key of $v_2$. Vehicle $v_3$ pretends to be $v_2$ and answers $v_1$ with $v_3$ public key instead of $v_2$. Then $v_3$ concurrently asks $v_2$ for its public key. Vehicle $v_1$ is fooled to establish a private key with $v_3$ instead of $v_2$, and $v_2$ is fooled to establish a private key with $v_3$ instead of $v_1$. Vehicle $v_3$ conveys messages from $v_1$ to $v_2$ and back, decrypting and re-encrypting with the appropriate established keys. In this way, $v_3$ can find the appropriate moment to change information and cause hazardous actions to $v_1$ and $v_2$.

\medskip\noindent\textbf{Our contribution.} Our work demonstrates the utility of out-of-band identification~\cite{hal,repo} using a coupled public key and fixed verifiable attributes. The certified attributes may be visually verified by a camera, microphone, wireless transceiver fingerprint identification~\cite{fing}, and other sensing devices. We ensure the countermeasures against the MitM attack in two sequential and explicit rounds of communication.

\begin{itemize}
\item \textit{Twofold authentication:} We propose a solution that employs a fixed attribute based certification mechanism to correctly identify the neighboring vehicles. Visual identification~\cite{newly} implies more robust authentication of the transmission source in comparison with the signal noise and/or transceiver fingerprint verification. Our solution relies on the verification that the public key was originated by the CA, and that the public key belongs to the vehicle with the coupled signed attributes. The periodic licensing routine can serve as an important ingredient of our protocol.


\item  \textit{Periodic certificate restore:} Our method has the benefit of interacting with the CA only during preprocessing stages, rather than during the real-time secret session key establishment procedure. Given such certified public key and vehicle attributes, the protocol establishes a secret session key with neighboring authenticated vehicles using only two communication rounds. 

 \item \textit{Adaptation:} The proposed approach can be integrated with the existing authentication protocols without beaching the respective security claims in these existing protocols. Therefore, the security claims are strengthened while adapting the proposed approach with the proven authentication protocols.

 \item \textit{Verification:} The proposed approach satisfies the secrecy and authentication properties. These security claims have been verified using an extended security analysis in Spi calculus.
\end{itemize}


\medskip\noindent\textbf{Related Work.}
In this section, we illustrate the related work, concerning vehicle network threats, state of the art for mitigating MitM attacks. Then we describe existing entity authentication schemes, and in particular, the utility of out-of-band communication for the authentication purposes.

\medskip\noindent\textit{Vehicle networks threats.}
An autonomous wireless connection among vehicles imposes serious security threats such as eavesdropping~\cite{rivest}, identity spoofing~\cite{spo,spoo}, sybil attack~\cite{sybi}, wormhole attack~\cite{wor}, replay attack~\cite{ply}, message content tampering~\cite{jean}, impersonation~\cite{impersonation}, denial of service attack (DoS)~\cite{dose} and Man-in-the-Middle attack~\cite{hbx}. In~\cite{baali} an anti-spoofing scheme based on Mutual Egress Filtering (MEF) using a compressed Access Control List (ACL) over border routers is presented. Furthermore, ~\cite{avtar,metrry} presents a survey of security challenges in Cognitive Radio Networks (CRN) with respect to exogenous/jamming, intruding, greedy attackers and crucial routing metrics.

\medskip\noindent\textit{Mitigating Man-in-the-Middle attacks.}
Global System for Mobile Communication (GSM) is one of the most popular standards in cellular network infrastructure. Unfortunately, it uses only one sided authentication between the mobile station and the coupled base station~\cite{dude}. The Universal Mobile Telecommunication Standard (UMTS) improves over the security loopholes in GSM. It includes a mutual authentication and integrity protection mechanism but is still vulnerable to MitM attacks~\cite{uts}.

MitM and DoS attack analysis for Session Initiation Protocol (SIP) is shown in~\cite{mim}, using a triangle communication model between SIP user agent and server. This work presents an analysis on the attack possibility, but does not offer any solution to the problem in hand. The interconnection between 3G and wireless LAN is vulnerable to MitM attacks by influencing the gateway nodes~\cite{3gwlan}. According to~\cite{dh}, mobile hosts and the base station share a secret cryptographic function and mutually raises a challenge-response string, prior to employing the original Diffie-Hellman key exchange scheme~\cite{direction}. Thus, the mobile host replies with a cryptographic response and Subscriber Station Identifier (SSI) to a base station, but it does not verify any of the unchangeable attributes of the intended subscriber. This way a base station, capable of verifying a unique SSI connection, may not confirm the authentic owner of the SSI connection. Furthermore, position-based routing schemes for vehicular networks~\cite{telecom,reusespace} would play a crucial role in a reliable secure communication. It may further be extended to energy saving such as a delay tolerant routing approach for vehicle networks~\cite{greendelay} that allows a delay bounded delivery by combining a carry-and-forward mechanism with the replication mechanism. Dynamic backpressure-based routing protocol for Delay Tolerant Networks (DTNs) is given in ~\cite{toldelay,disruptol,dvir}. Accordingly, routing decisions are made on a per-packet basis using queue logs, random walk and data packet scheduling as opposed to static end-to-end routes. Similarly, a stability based routing scheme to synergise noise ratio, distance and velocity into a routing decision is given in~\cite{tability}. A reliable multicast protocol for lossy wireless networks using opportunistic routing with random linear network coding~\cite{opprtu,codepipi,opormatter} and a genetic algorithm based approach is presented in~\cite{yenfmq}.

\medskip\noindent\textit{Entity authentication.}
There has been a great research activity in the scope of cryptographic solutions~\cite{two} for the entity authentication.
A security scheme for sensor networks, called TESLA has been proposed in~\cite{fifteen}. TESLA is based on delayed authentication with self-authenticating key chains. TESLA yields a time consuming authentication mechanism (as the messages are received on a timeline can be authenticated only after receiving the immediate next message over the same timeline). Although, chances are less, but a MitM can still intercept through weak hash collisions and fake delayed key.
An improvement TESLA++ has been suggested in~\cite{sixteen}, as an adapted variation of delayed authentication. A combination of TESLA++ and digital signature provides Denial of Service (DoS) attack resilience and non-repudiation respectively. The drawback with this approach is that the message digest and corresponding message (with self-authenticating key) are transmitted separately to the receiver. Thus, MitM may step in, as it does not follow the fixed attribute based verification. Furthermore, another scheme for anomaly-detection and attack trace back for encrypted protocols~\cite{features} and IP spoofing trace back is given in~\cite{fore}.

Raya and Haubaux~\cite{raya,mraya} proposed that each vehicle contains a set of anonymous public/private key pairs, while these public keys have been certified by CA. The certificates are short lived, and therefore, need to be confirmed with a Certificate Revocation List (CRL) before the use. The drawback with this approach is that roadside infrastructure is required to provide the most updated CRL. A MitM attack resistant key agreement technique for the peer to peer wireless network is suggested in~\cite{seventeen} where the primary mutual authentication is done before the original Diffie-Hellman key exchange. This primary authentication step could be a secret digest comparison, e.g., through visual or verbal contact, distance bounding or integrity codes. However, a MitM can intercept because of the proximity awareness, visual and verbal signals, computed by the device and verified by the user; while in our case it is already certified by the CA and then user verifies it again. Moreover, the authors in~\cite{softsecure} presented an incentive mechanism for peer-to-peer networks in order to encourage user cooperation as opposed to selfish behaviour and another Software Defined Network (SDN) security in~\cite{haali}.
The secure communication scheme, in~\cite{dblp25} is an enhancement over the Raya and Haubaux scheme, in that a certified public key is exchanged and further used to set up a secret session key as well as a group key. Here, the attacker can pretend to be some other vehicle, by replaying the certificates and there currently exists no other means to verify that this vehicle is not the actual owner of the certificate. In addition~\cite{bengou,yinli} presents a secure (i.e. proactive secret sharing scheme) and privacy preserving (identity blinding) key management scheme resilient to time and location based mobile attacks proposed for the m-healthecare social networks~\cite{health}.

\medskip\noindent\textit{Out-of-band channel authentication.}
There have been great efforts to utilize various auxiliary out-of-band channels for entity authentication. The notion of pre-shared secret over a limited contact channel was first raised in~\cite{duckling}.

The pre-authentication channel is a limited scope channel to share limited information but it inherits the same vulnerability that a wireless channel has. In this scheme, there may be cases when a vehicle authenticates the sender but is not able to verify the specific identity traits. In our scheme we do it in reverse, first the certified attribute verification over the wireless channel and then the static attribute verification over the out-of-band channel of communication.

A method shown, in~\cite{smart,shakewell}, suggests that a common movement pattern can help to mutually authenticate two individual wireless devices driven by a single user. In~\cite{strangers}, a pre-authentication phase is used to verify the identity, before the original public key is exchanged and confirmed over the insecure wireless channel. Another work, in~\cite{sib}, presents a visual out-of-band channel. A device can display a two dimensional barcode that encodes the commitment data, hence, a camera equipped device can receive and confirm this commitment data with the available public key. Unfortunately the attacker can still capture and/or fabricate the visible commitment data, as it is not coupled with the public key. The approach in~\cite{loudc} is based on acoustic signals, using audio-visual and audio-audio channels to verify the commitment data. In the audio-visual scheme, a digest of the public key is exchanged by vocalizing the sentence and comparing with a display on the other device, while the audio-audio scheme, compares vocalized sentences on both devices. In a recent work~\cite{blink}, Light Emitting Diode (LED) blinks and the time gap between those blinks has been used to convey the digest on the public key. Also, a combination of an audio-visual and an out-of-band channel has been proposed in~\cite{bplnk}, that uses beeps (audio) and LED blinks (visual) in a combination to convey the commitment data. The proposed method is less effective because the public key and the out-of-band information is not certified therefore MitM can record the out-of-band information and replay it. The approach in~\cite{mayrhofer} suggested the use of spatial reference authentication such as correlating latency with the distance which can be faked by a MitM attacker. In particular, all identification techniques presented in~\cite{mayrhofer} are not coupled with the public key in a signed form, therefore, allows an MitM attacker to penetrate.


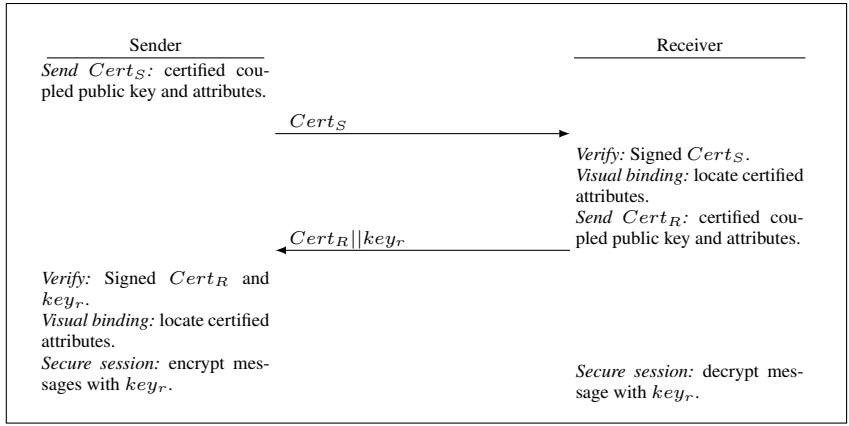
\begin{figure}
\captionsetup{belowskip=2pt,aboveskip=8pt}
\centering
\scriptsize
\begin{tikzpicture}[show background rectangle,inner frame sep=.3cm]
\matrix (m)[matrix of nodes, column  sep=1mm,row  sep=1mm, nodes={draw=none, anchor=center,text depth=0pt} ]{
Sender & & Receiver\\[-1mm]
\begin{minipage}{3cm}
\textit{Send $Cert_S$:} certified coupled public key and attributes.
\end{minipage}   & & \\ [1mm]
 &\begin{minipage}{3.5cm}
 $Cert_{S}$
 \end{minipage}  & \\
& & \begin{minipage}{3cm}
\textit{Verify:} Signed $Cert_{S}$.

\textit{Visual binding:} locate certified attributes.

\textit{Send $Cert_R$:} certified coupled public key and attributes.

\end{minipage} \\ [1mm]
 & \begin{minipage}{3.5cm}
 $Cert_{R}||key_{r}$
 \end{minipage} & \\[1mm]
 \begin{minipage}{3cm}
 \textit{Verify:} Signed $Cert_{R}$ and $key_r$.

 \textit{Visual binding:} locate certified attributes.

 \textit{Secure session:} encrypt messages with $key_r$.
 \end{minipage} & \\ [1mm]
& &  \begin{minipage}{3cm}
\textit{Secure session:} decrypt message with $key_r$.
 \end{minipage} & \\ [1cm]
};

\draw[shorten <=-1cm,shorten >=-1cm] (m-1-1.south east)--(m-1-1.south west);
\draw[shorten <=-1cm,shorten >=-1cm] (m-1-3.south east)--(m-1-3.south west);
\draw[shorten <=-1mm,shorten >=-1mm,-latex] (m-3-2.south west)--(m-3-2.south east);
\draw[shorten <=-1mm,shorten >=-1mm,-latex] (m-5-2.south east)--(m-5-2.south west);
\end{tikzpicture}
\caption{The proposed protocol.}
\label{fig:originalMP}
\end{figure}

Section \ref{sect:prot} illustrates the system settings and a detailed description of the proposed authentication scheme. Next, in Section \ref{sect:AKE}, we discuss properties of our proposition in relation to the security provided by other key establishment protocols and the transport layer security handshake with certified attributes. Section \ref{sect:correct}, demonstrate a high level security analysis along with a formal correctness sketch using Spi calculus. The last Section \ref{sect:conclusion} concludes the discussion.

\section{Out-of-band Sense-able Certified Attributes for Mitigating Man-in-the Middle Attacks} \label{sect:prot}
The proposed approach is specifically designed and ready to use for the recently customized vehicles with following configurations.

\medskip\noindent\textbf{System Settings.}

\medskip\noindent\textit{Customized standards and hardware for vehicles.}
The vehicles are assumed to be equipped with Electronic Control Units (ECUs), sensors, actuators~\cite{four} and the wireless transceiver that support the DSRC (Dedicated Short Range Communication) standard~\cite{cch,dsrc}. These ECU's are interconnected over a shared bus to trigger a collaborative decision on some safety critical events. Furthermore, a wireless gateway is installed to connect the in-vehicle network with the external network or device for the diagnosis purposes. The in-vehicle network can be divided into the controller area network (CAN), local interconnect network (LIN), and media oriented system (MOST)~\cite{dblp24} based on the technical configurations and the application requirements. These embedded devices enable facilities such as automatic door locking, collision warnings, automatic brake system, reporting road conditions, rain and dark detection and communication with the surrounding road infrastructure. Therefore, vehicles must be equipped with the fundamental communication capabilities as per the vehicular communication standards mentioned above. Our protocol would provide a secure communication on top of the available and standardized communication schemes in these customized vehicles.

\medskip\noindent\textit{Registration and identity certification.}
In addition, with the technical configurations that a vehicle must be equipped with, a trusted third party is also required periodically for the successful execution of the proposed approach in this paper. Currently, every vehicle is periodically registered with its national or regional transportation authority, which allocates a unique identifier to the vehicle with an expiration date which usually is the next required inspection date. In some regions of the United States and Europe, registration authorities have made substantial progress toward electronically identifying vehicles and machine readable driving license. According to the state of the art these registration authorities would assign a public/private key pair to the inspected vehicles for a secret information exchange. However, our protocol would be secure against the possible attacks even if vehicles are pre-assigned with the certificates.

We suggest mitigating MitM attacks by coupling out-of-(the wireless)-band verifiable attributes (see Figure ~\ref{fig:originalMP}). Vehicles are authenticated using digitally signed certificates and out-of-band verifiable attributes. For example, these attributes may include visual information that can be verified by input from a camera when there exists line-of-sight including the identification of the driving license number, brand, color and texture, and even the driver's face if the owner wants to restrict the drivers that may drive the vehicle. Other attributes may be verified by other sensing devices, such as a microphone for noise.

\begin{figure}[t]
  \begin{center}
    \includegraphics[width=10cm]{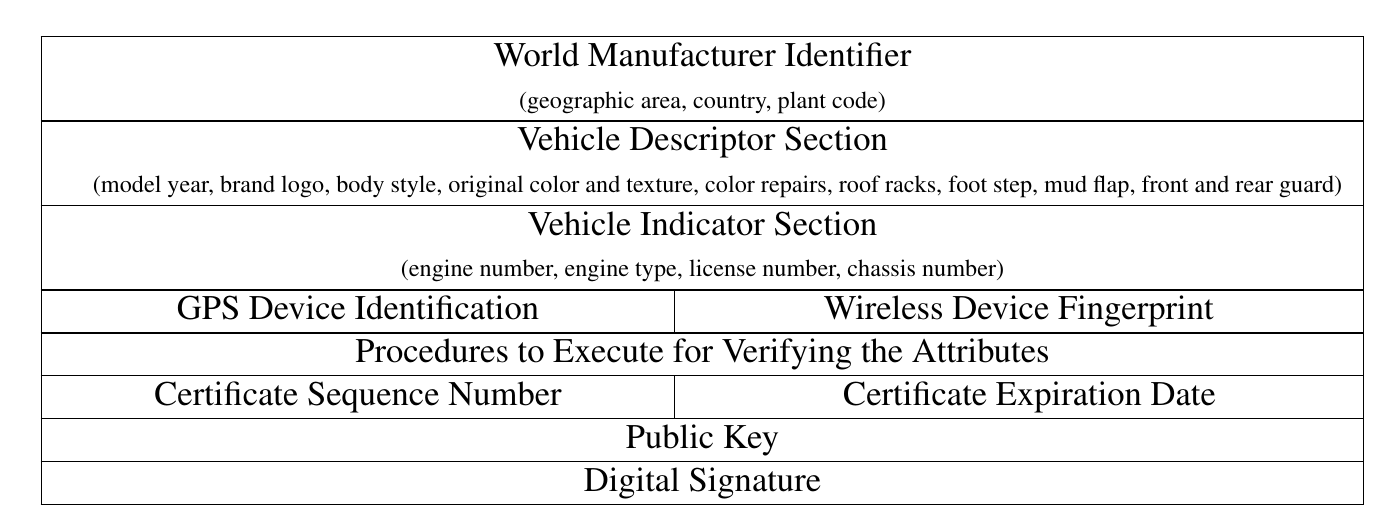}
  \end{center}
  \caption{Certificate structure.}
  \label{figure:certstruc}
\end{figure}

The proposed protocol can be pinpointed as follows:
\begin{itemize}
    \item Initially, vehicles must preprocess a uniform cipher suite and a unique certificate from a CA.
    \item Communication starts with exchanging a digitally signed certificate that is a commitment over the certified attributes and coupled public key.
    \item The present protocol utilizes an indirect binding over the commitment data, and the shared secret session key.
    \item Two rounds of session key negotiation must ensure the authenticity, secrecy of the origin, and message contents, respectively (the proposed protocol with improvements against impersonation attacks, see Section \ref{sect:AKE}).
    \item Commitment data is assumed to be hashed using a collision free and second preimage resistant function.
    \item The protocol is interactive, while enabling the mutual authentication, session key establishment and subsequent session interaction in a single protocol run.
    \item The protocol ensures perfect forward secrecy (protocol with improvements for FS, see Section \ref{sect:AKE}).
\end{itemize}

\begin{figure}[t]
\captionsetup{width=0.90\textwidth,belowskip=0pt,aboveskip=14pt}
\centering
\begin{tikzpicture}[show background rectangle,inner frame sep=2mm]
\node [mybox] (box){%
    \begin{minipage}{0.90\textwidth}
    \begin{enumerate}

  \item Sender $S$ sends the certificate $Cert_S=Attribute_{S}+Public \: key_{S}||Sign_{CA}(H(Attribute_{S}+Public \: key_{S}))$ to a neighbor $R$.

  \item Receiver $R$ confirms the certificate $Cert_S$ authenticity as described in 2.(a) and then responds as detailed in 2.(b):
\begin{enumerate}

  \item $R$ verifies the digital signature using the CA public key $PK_{CA}$ and verifies $Attribute_S$ using out-of-band channels.

  \item $R$ responds with the certificate $Cert_R=Attribute_{R}+Public \: key_{R}||Sign_{CA}(H(Attribute_{R}+Public \: key_{R}))$. Also appends a random string $key_r$ and certificate sequence number $Sequence \: Number_S$ encrypted with $Public \: key_S$ and $SK_{R}$, i.e. $E_{Public \: key_{S}}(key_r+Sequence \: Number_S)||E_{Public \: key_S}(E_{SK_R}(H(key_r+Sequence \: Number_S)))$.
  \end{enumerate}

  \item Sender $S$ confirms the certificate $Cert_R$ authenticity as described in 3.(a) and then responds as detailed in 3.(b):
\begin{enumerate}

  \item $S$ verifies the digital signature using the CA public key $PK_{CA}$ and verifies $Attribute_R$ using out-of-band channels.

  \item $S$ decrypts the secret session key and certificate sequence number concatenated with the digital signature by using own secret key $SK_S$, i.e. $D_{SK_S}[E_{Public \: key_{S}}(key_r+Sequence \: Number_S)]$ resulting into $key_r$. Also the digital signature of $R$ is verified using $SK_S$ and $Public \: key_R$ respectively, i.e. $D_{SK_S}(D_{Public \: key_R}(H(key_r+Sequence \: Number_S)))$ that results into $H(key_r+Sequence \: Number_S)$. Now the hashing algorithm $H$ is applied with $key_r+Sequence \: Number_S$ and then compared with the hashed string $H(key_r+Sequence \: Number_S)$ produced from the digital signature. If both hash strings are similar and the symmetric padded zero composition $key_r+Sequence \: Number_S$ is valid, then $key_r$ is accepted as a valid session key.

  \end{enumerate}

\item Sender and receiver exchange encrypted messages using $key_r$ as a shared secret key for $S$ and $R$.

\end{enumerate}
    \end{minipage}
};
\end{tikzpicture}%
 \caption{Session key establishment in two rounds.}

 \label{fig:dcv}
\end{figure}

In the proposed protocol, vehicles carry a digitally signed certificate $Cert$ from CA (see Figure \ref{figure:certstruc}) for a possible structure of such a certificate. The pseudo-code description of the secret key establishment procedure appears in Figure \ref{fig:dcv} and notations in Table ~\ref{table:notations}. Our protocol does not require any communication with the CA or the road side units, while actually authenticating vehicles on the move. The only interaction with the CA is during a preprocessing stage, which is mandatory to possess a certificate. The certificate holds a public-key and unchangeable (or rarely changeable) attributes of the vehicle signed by the CA. These out-of-band sense-able vehicular attributes should be sensed by other vehicles and checked in real-time. Note that the procedure to check these vehicular attributes may be given as part of the certified information. Our protocol is a viable solution to combat the MitM attacks, as it utilizes a separate sense-able out-of-band channel to authenticate the unchanged vehicular attributes. The certificate can be updated and restored on each periodical inspection or in the rare case of an attribute change. Therefore, the proposed approach saves time and communication overhead in the authentication process. In addition, it avoids the frequent CA communication bottleneck and is suitable for the emergency and safety critical applications. A detailed description of the solution appears in the next section.

\begin{table}[t]
\begin{center}
\begin{spacing}{0.85}
            \begin{tabular}{|l|l||l|l|}
            \hline
               {\scriptsize $S$} & {\scriptsize Sender} & {\scriptsize $R$} & {\scriptsize Receiver} \\ \hline

               {\scriptsize $Cert_S$} & {\scriptsize Certificate of sender} & {\scriptsize $Cert_R$} & {\scriptsize Certificate of receiver} \\ \hline

               {\scriptsize $PK_{CA}$} & {\scriptsize Public key of $CA$} & {\scriptsize $SK_{CA}$} & {\scriptsize Secret key of $CA$} \\ \hline

               {\scriptsize $PK_S$} & {\scriptsize Public key of $S$} & {\scriptsize $PK_R$} & {\scriptsize Public key of $R$} \\ \hline

               {\scriptsize $SK_S$} & {\scriptsize Secret key of $S$} & {\scriptsize $SK_R$} & {\scriptsize Secret key of $R$} \\ \hline

               {\scriptsize $Attribute_S$} & {\scriptsize Static attributes of $S$} & {\scriptsize $Attribute_R$} & {\scriptsize Static attributes of $R$} \\ \hline

               {\scriptsize $Sequence \: Number_S$} & {\scriptsize Sequence number of $S$} & {\scriptsize $Sequence \: Number_R$} & {\scriptsize Sequence number of $R$} \\ \hline

               {\scriptsize $H$} & {\scriptsize Hash function} & {\scriptsize $key_{r}$} & {\scriptsize Session key} \\ \hline

               {\scriptsize $||$} & {\scriptsize String concatenation} & {\scriptsize $+$} & {\scriptsize symmetric bit padding} \\ \hline

               {\scriptsize $E_{PK}$} & {\scriptsize Encryption with PK} & {\scriptsize $D_{PK}$} & {\scriptsize Decryption with PK} \\ \hline

               {\scriptsize $E_{SK}$} & {\scriptsize Encryption with SK} & {\scriptsize $D_{SK}$} & {\scriptsize Decryption with SK} \\ \hline

               {\scriptsize $v$} & {\scriptsize Vehicle} & {\scriptsize $l$} & {\scriptsize License number} \\ \hline
            \end{tabular}
\caption{Notations.}
\label{table:notations}
\end{spacing}
\end{center}
\end{table}

We assume that the CA established a certificate in the form of $Attribute_{S}+Public \: key_{S} \: || \: Sign_{CA}(H(Attribute_{S}+Public \: key_{S}))$ for each party. These certificates are used to establish a (randomly chosen) shared key, $key_r$. The shared key $key_{r}$ can then be used to communicate encrypted information from the sender to the receiver and back. One way to do this is to use the $key_r$ as a seed for producing the same pseudo-random sequence by both the sender and the receiver. Then XOR-ing the actual sensitive information to be communicated with the bits of the obtained pseudo-random sequence. Next, we describe in detail the involved entities and their part in the procedure for establishing a session key.

\medskip\noindent\textbf{Certificate Authority.}
The list of CAs with their public keys $PK_{CA}$ may be supplied as an integral part of the transceiver system of the vehicle, similar to the way browsers are equipped with a list of CAs public keys.
Only registered vehicles are allowed to communicate on the road. Digital signatures $Sign_{CA}(H(Attribute_{sender}+Public $\:$ key_{sender}))$ represent the hash of the public key and attributes encrypted with the CA secret key $SK_{CA}$. The digital certificate works as an approval over the public key and the out-of-band verifiable attributes of the vehicle. The CA can update or renew a certificate, upon a need or when the current certificate expires.

\medskip\noindent\textbf{Vehicular Attributes.}
A vehicle incorporates various sensors to capture useful primitives from the neighborhood. Each vehicle is bound to a set of primitives yielding a unique identity to that vehicle. Vehicle identity encloses a tuple comprised of attributes such as license number, public key, distinct visual attributes and other out-of-band sense-able attributes, extending the basic set of attributes required according to ISO 3779 and 3780 standard~\cite{iso}.
The idea behind using out-of-band attributes such as vision and position is to simulate the human perception in real world along with the digital signatures to confirm the identity. These out-of-band sense-able attributes are captured through customized device connections such as camera, microphone, cellular communication and satellite (GPS system). In addition, we suggest identifying the wireless communication itself, rather than the contents sent by the wireless communication, this is done by the certified transceiver fingerprints. Thus, the transceiver must be removed from the original vehicle and possibly be reinstalled in the attacker's vehicle to launch the attack.
Verifying each of the attributes by out-of-band channel implies a certain trust level in the identity of the communicating party, which in turn implies the possible actions taken based on the received information from the partially or fully authenticated communicating party.
Thus, a vehicle can perceive the surroundings from driver's perspective using vision with a sense of texture, acoustic signals, and the digital certificate. A combination of these primitives is different for every vehicle, i.e., license number, outlook of the vehicle including specific equipment, specific visual marks such as specific color repair marks, manufacturer's logo and/or engine acoustics classification signals.

\section{Key Exchange Protocols with Out-of-Band Sense-able Attributes Authentication} \label{sect:AKE}
Many two-party Authenticated Key Exchange protocols (AKE)\cite{Law98anefficient,DBLP:conf/crypto/Krawczyk03,Lamacchia06strongersecurity,nxplus,DBLP:conf/pkc/LauterM06} which allow two parties to authenticate each other and to establish a secret key via a public communication channel and three-party AKE protocols~\cite{threparty} have been proposed over the past years addressing various adversary models and possible attacks. There exists one-round protocol that ensures weak forward secrecy~\cite{krawczyk} that is providing \textit{Forward Secrecy} only when the adversary is not active in the session. These one round protocols are based on a simultaneous interaction between the sender and receiver. In their work, they also prove the impossibility of establishing a strong forward security in one round. However, one-way protocol with strong secrecy exists in~\cite{oneround,rabin,neito}. They have assumed that the ephemeral secret keys are exchanged between the peer parties while the adversary is not allowed to access the ephemeral secret key. Informally, as it is stated in \cite{DBLP:conf/crypto/Krawczyk03}, AKE protocols should guarantee the following requirements:
\textit{Authentication} -- each party identifies its peer within the session;
\textit{Consistency} -- if two honest parties, $A$ and $B$, establish a common session key $K$, then
$A$ believes it communicates with $B$ while $B$ believes it communicates with $A$;
\textit{Secrecy} -- if a session is established between two honest peers, then no adversary
should learn any information about the resultant session key.

Usually the above requirements are more formally described by detailed scenarios that involves resistance to the following attacks:
\textit{Basic Key Exchange (KE) security} is defined through the KE experiment in which an adversary that controls a communication channel should not be able to distinguish the session key established between parties from a random value.
\textit{Forward Secrecy (FS)} property guarantees that a session key derived from a set of long-term public and private keys will not be compromised if one of the (long-term) private keys is compromised in the future.
Therefore, an adversary who corrupted one of the parties (learns the long-term secret key), should not be able to learn session keys of past sessions executed by that party.
\textit{Known Session Key Attack resilience} provides that an adversary who learns a session key  should be unable to learn other session keys.

Additionally, authentication in AKE protocols implies resistance to various misidentification threats:
\textit{Unknown Key-Share Attacks resilience} prevents an adversary to cause the situation whereby a party (say $A$), after the protocol completion believes she shares a key with $B$, and although this is in fact the case, $B$ mistakenly believes the key is shared with a party $E$ (other than $A$).
\textit{Key Compromise Impersonation (KCI) resilience} provides that an adversary who learns a long-term secret key of some party (say $A$) should be unable to share a session key with $A$ by impersonation as the other party to $A$, although obviously it can impersonate $A$ to any other party.
\textit{Extended Key Compromise Impersonation (E-KCI) resilience.} In regular AKE protocols, parties use additional random parameters known as \textit{ephemeral} keys, for example, ephemeral Diffie-Hellman keys coined for the purpose of session initialization. An adversary who learns both a long-term secret key, and an ephemeral key of some party (say $A$), should be unable to share a session key with $A$ by impersonation as another party to $A$.
\textit{Ephemeral Key Compromise Impersonation (ECI) resilience.}  An adversary who learns only an ephemeral key of some party (say $A$), should be unable to share a session key with $A$ by impersonation as another party to $A$.

In this paper, we focus on specific AKE scenarios for securing the communication of vehicles via out-of-band sensible attributes.
We assume that:
\begin{enumerate}
   \item a sender and a recipient use specialized devices for recognizing out-of-band sensible attributes.
	\item these devices can precisely pick the peer vehicle, and can accompany a regular (say radio communication) channel. 
    \item the out-of-band sensible attributes can identify a vehicle uniquely.
\end{enumerate}


\begin{figure}
\captionsetup{width=0.40\textwidth,belowskip=4pt,aboveskip=10pt}
 	\centering
 		\includegraphics[width=0.80\textwidth]{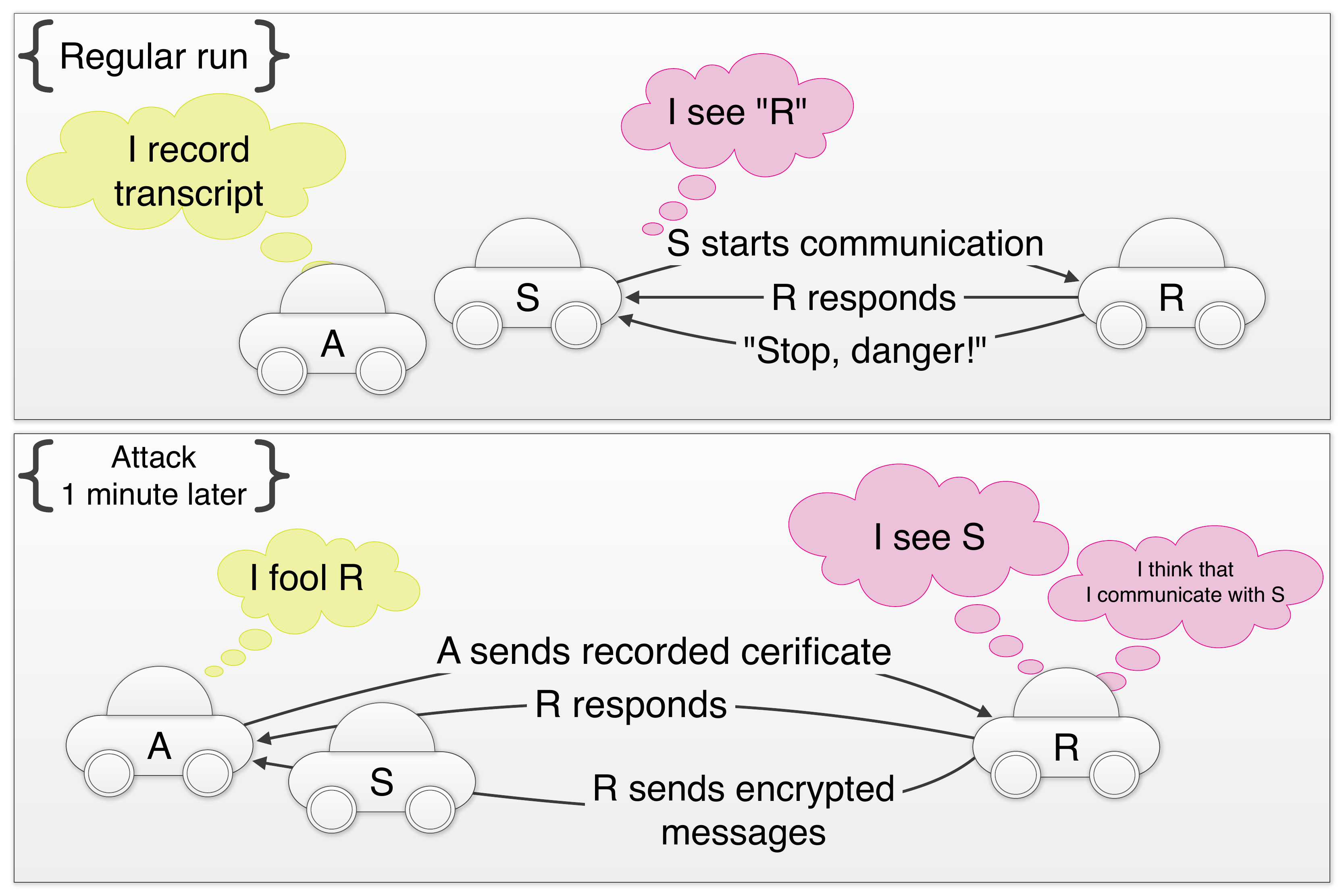}
 	\caption{Repetition attack - version 1.}
 	\label{fig:attack1}
 \end{figure}

 If the above mentioned assumptions do not hold, then the protocol in Figure \ref{fig:dcv} can be a subject of impersonation repetition attacks and do not fulfill FS feature, as it is outlined below.

\medskip\noindent\textbf{Impersonation Repetition attack - version 1.} Any adversary $A$ that is within the radio range  of a sender $S$ with $Attribute_{S}$ and a recipient $R$ with $Attribute_{R}$, that once recorded a valid transcript and the certificate of $S$, can initialize 
future communication from $S$. Although $A$ cannot decipher responses from $R$, the attack could be used to make $R$ thinking that $S$ wants to communicate. Moreover, $R$ can use such an initialized session to send some valid but unwanted messages to $S$ (see Figure \ref{fig:attack1}).

\begin{figure}[b!]
\captionsetup{width=0.40\textwidth,belowskip=4pt,aboveskip=10pt}
 	\centering
 		\includegraphics[width=0.80\textwidth]{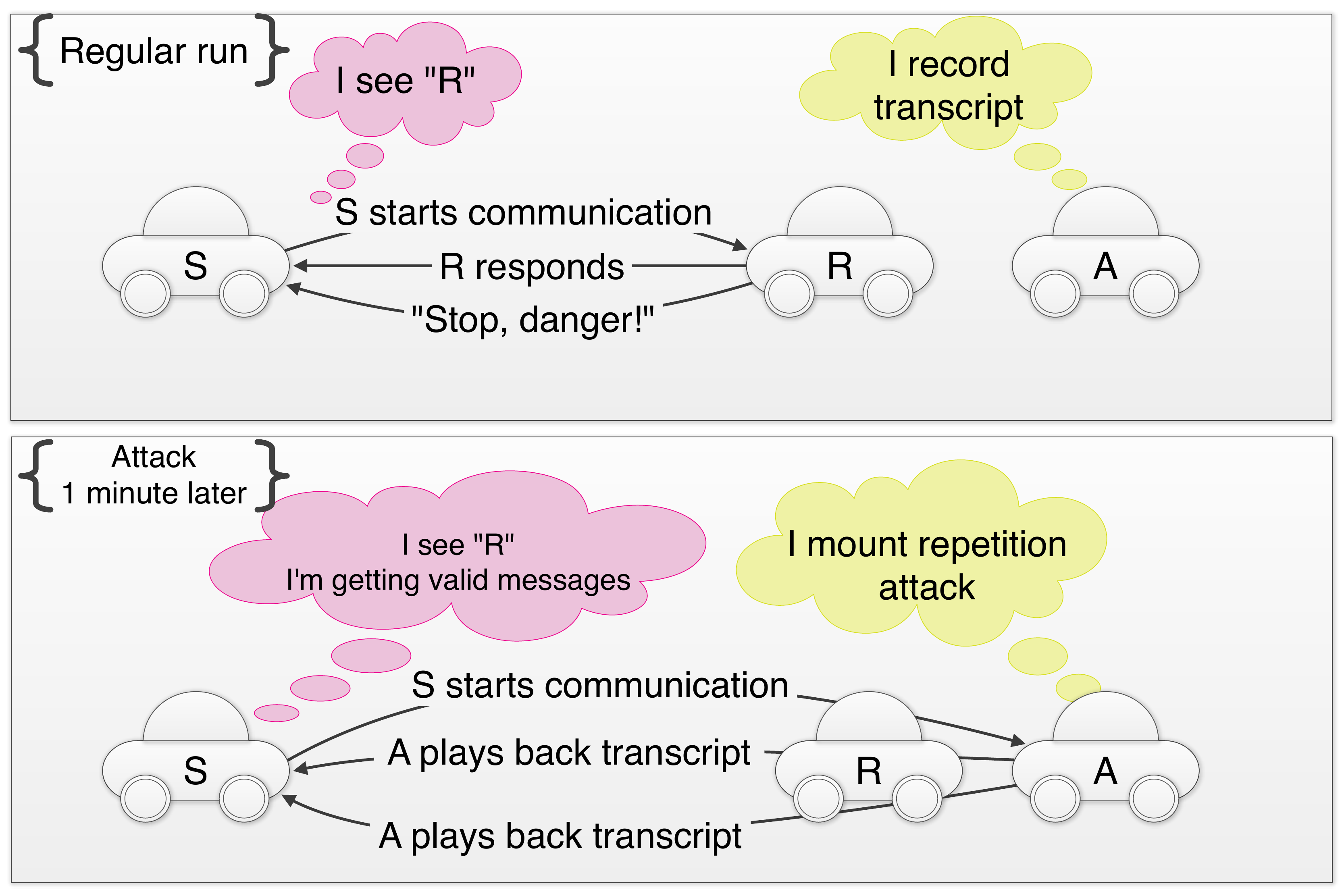}
 	\caption{Repetition attack - version 2.}
 	\label{fig:attack2}
 \end{figure}

\medskip\noindent\textbf{Impersonation Repetition attack - version 2.} An adversary $A$ that once recorded a valid transcript between a sender $S$ with $Attribute_{S}$ and a recipient $R$ with $Attribute_{R}$ can simulate future answers (steps 2a, 2b in Figure \ref{fig:dcv}) for the same recipient $R$ (or for any other recipients $R'$ - that has similar attributes $Attribute_{R}$) challenged by $S$. Adversary $A$ simply sends back messages previously recorded in steps 2a, 2b (see Figure \ref{fig:attack2}). Thus, after $S$ finishes protocol in accepting state, it thinks it partnered with the intended $R$, and starts to decrypt subsequent messages encrypted with the established key.
Although, in this repetition attack, $A$ does not learn the session key, after acquiring the first message from $S$, the adversary $A$ can send back previously recorded answers from $R$ to $S$, finishing protocol. Subsequently, $A$ can continue with sending previously recorded cipher texts encrypted with the previous session key. Such cipher texts would be accepted as valid, and decrypted by $S$.  If the protocol run is aimed only for the authentication purposes (peers do not want to communicate further, which we do not consider here), then the attack itself is a serious threat, e.g., the case where $S$ is a police car that monitors the speed of other cars and wants to identify the recipient.

\textit{Improvements Against Impersonation Attacks.} In the case of the proposed protocol, we can simply protect against impersonation attack version 1 in the following way: a sender $S$ encrypts an acknowledgment of the second message it receives from $R$ with the session key and sends at the beginning of the transmission through the encrypted channel. For the protection against the impersonation attack version 2, a sender $S$ sends (in the first step) a concatenation $Cert_S|Nonce_S$ to $R$, where $Nonce_S$ is a unique random challenge coined for that session by $S$. Then the cryptograms answered by $R$ in the second step should include the same $Nounce_S$, which subsequently should be verified by $S$.

\begin{figure}[b!]
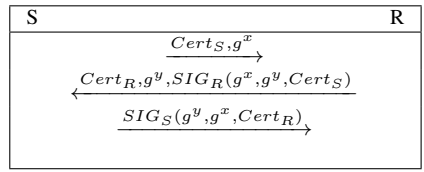

{\small
\begin{center}
\begin{tabular}{|c c c|}
\hline
 S &  & R \\
\hline
& $\xrightarrow{Cert_S, g^x}$ & \\
& $\xleftarrow{Cert_R, g^y, SIG_R(g^x,g^y, Cert_S)}$ & \\
& $\xrightarrow{SIG_S(g^y,g^x, Cert_R)}$ & \\
& & \\
\hline
\end{tabular}
\end{center}
\caption{ISO KE adopted to the proposed certificates. \label{fig:ISO-KE}}
}
\end{figure}

\textit{Forward Secrecy} (FS):  This is the protection of past session keys in spite of the compromise of long-term secrets. If the attacker somehow learns the long-term secret information held by a party (the party is controlled by the attacker, and referred to as corrupted), it is required that session keys, produced (and erased from memory) before the party corruption happened, will remain secure (i.e. no information on these keys should be learned by the attacker). Obviously our protocol does not fulfill FS. If the attacker records transcripts and then corrupts the party $S$ (got its private keys), then the previous session keys $key_r$ are exposed and transcripts can be deciphered.
\textit{Improvements for FS.} We can improve our protocol for FS by setting: $Nounce_S=g^{\alpha}$, responded $key_r=g^{\beta}$, for some random ephemeral keys $\alpha$, and $\beta$. Then the session key would be derived from the value $g^{\alpha\beta}$ and computed independently on both sides.

Obviously one can also utilize some three-round protocols, instead of our two rounds protocol, protocols previously discussed in literature, that
do not require a predefined knowledge of peers identity. The idea of out-of-band sense-able attributes can be incorporated into them without undermining their security. The first straightforward choice would be ISO KE protocol, described in \cite{ISO/IEC-IS-9798-3}, and mentioned among other protocols in \cite{DBLP:conf/crypto/Krawczyk03}. Figure \ref{fig:ISO-KE} presents the protocol, where $Cert_S$, and $Cert_R$  are certificates proposed in this paper. In the protocol, any vehicle receiving the certificate can immediately validate the certificate by means of the CA public key, and out-of-band visible attributes. They also validate received signatures and proceed only if the validation is correct. The established session key $K_S$, is derived from $g^{xy}$. Note that this protocol does not support identity hiding, as certificates are transferred in plain texts.

\begin{figure}
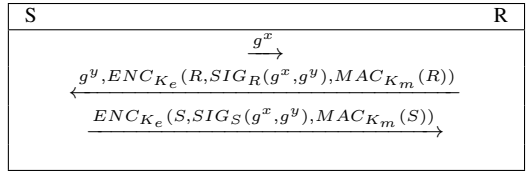

{\small
\begin{center}
\begin{tabular}{|c c c|}
\hline
S &  & R\\
\hline
& $\xrightarrow{g^x}$ & \\
& $\xleftarrow{g^y, ENC_{K_e}(R, SIG_R(g^x,g^y), MAC_{K_m}(R))}$ & \\
& $\xrightarrow{ENC_{K_e}(S, SIG_S(g^x,g^y), MAC_{K_m}(S))}$ & \\
&  & \\
\hline
\end{tabular}
\end{center}
\caption{SIGMA protocol adopted to the proposed certificates \label{fig:SIGMA}}
}
\end{figure}

If we consider anonymity where the communicating entities must not be exposed to the non-communicating entities then the certificates should not be transferred as plain texts. The SIGMA protocol~\cite{DBLP:conf/crypto/Krawczyk03} for identity protection is based on a DH exchange authenticated with the digital signatures. A session key $K_S$, an encryption key $K_e$ and a message authentication key $K_m$ are derived from $g^{xy}$ ($K_S$, $K_e$, and $K_m$ keys must be computationally independent from each other), see Figure~\ref{fig:SIGMA}. Here, parties decrypt messages by the means of the key $K_e$,  validate certificates by the means of CA public key and verify the MACed identity. Each part independently proceeds only if both the decryption and validation are correct. However, this allows MitM attacks as the CA is not involved during the on-line process of key exchange. Moreover, the two parties in communication might not verify the mutual identity as if they communicate with the actual intended party or some other party holding valid certificate as well as identity (may be an adversary or an innocent identity misbinding). Therefore, proposed approach provides a coupling between the vehicle's physical identity and the authenticated communication over wireless channel.

If deniability property (that assures that transcript should not be regarded as a proof of interaction) is important, then we propose to adopt one of the protocols \cite{MRI2013,MCA2013}. However, in this case we should assume that parties private keys are discrete logarithms of corresponding public keys, and computations are performed in algebraic structures where the discrete logarithm problem (DLOG) is hard. Although deniable protocols from \cite{MRI2013,MCA2013} require four passes of messages, they were designed for machine readable travel documents - which in turn can be implemented on smart-cards. Therefore, we acknowledge that implementing them for vehicular communication can also be considered.

In addition, our protocol can be combined with the well-known existing authentication protocols, e.g., NAXOS~\cite{naxos}, NAXOS+~\cite{nxplus} that is proven to be secure in CK~\cite{ck} and eCK~\cite{naxos} security models. NAXOS assumes that sender and receiver have already exchanged the public key/certificate and requires additional two rounds for the ephemeral key exchange and session key establishment. However, the proposed protocol provides a certified visual binding in two explicit rounds of certificate exchange and does not interfere with the security claims of associated authentication protocol.

\begin{figure}[ht]
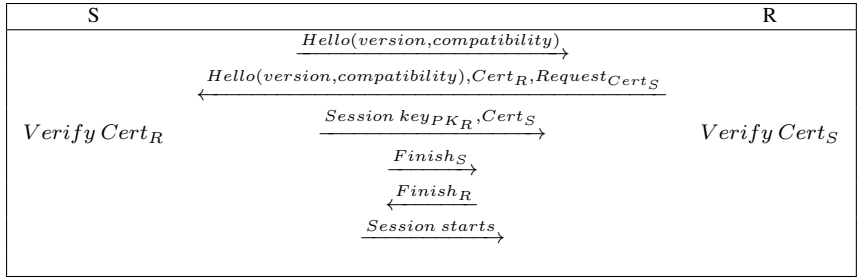

{\small
\begin{center}
\begin{tabular}{|c c c|}
\hline
S &  & R\\
\hline
& $\xrightarrow{Hello (version, compatibility)}$ & \\
& $\xleftarrow{Hello (version, compatibility), Cert_R, Request_{Cert_S}}$ & \\
$Verify \: Cert_R$ & $\xrightarrow{Session \: key_{PK_R}, Cert_S}$ & $Verify \: Cert_S$  \\
& $\xrightarrow{Finish_S}$ & \\
& $\xleftarrow{Finish_R}$ & \\
& $\xrightarrow{Session \: starts}$ & \\
& & \\
\hline
\end{tabular}
\end{center}
\caption{Transport Layer Security \label{fig:TLS}}
}
\end{figure}


\medskip\noindent\textbf{Transport Layer Security Handshakes with Certified Attributes.}
Our proposed approach adapts the security construction of the conventional Transport Layer Security (TLS) protocol as depicted in Figure~\ref{fig:TLS}. Accordingly, TLS mutual authentication is based on a certificate exchange between the sender and receiver. Apparently, our proposed approach inherits the certificate based security handshake framework from TLS protocol, additionally, the ability to verify the certified attributes through an auxiliary communication channel is the key contribution of our protocol. We propose a modified certificate structure, that certifies the coupling between visual static attributes and the public key of a vehicle. Therefore, the receiver verifies the integrity and authenticity of the certified and coupled public key and static attributes through an out-of-band communication channel. In addition, sender and receiver verify this coupling before switching on to a wireless communication channel. TLS handshakes are based on a pre-defined sequence of phases such as mutual authentication, random secret exchange and session key establishment. However, the handshake between sender $S$ and receiver $R$ starts by sending the supported range of cryptographic standards, also called as $Hello$ message. Moreover, the mutual authentication is accomplished through the CA signed certificates called a $Certificate \: Exchange$ message.

At first, $S$ forwards the certificate $Cert_S$ to $R$ which then verifies the CA signature on $Cert_S$ and the out-of-band sense-able fixed attributes of the sender, i.e., $Attribute_S$. Similarly, $S$ also verifies the CA signature on $Cert_R$ and the out-of-band sense-able fixed attributes $Attribute_R$. During the certificate exchange receiver $R$ generates a random string $key_r$ and forwards to $S$ along with the certificate $Cert_{R}$. The random string is encrypted with the intended receiver's certificate sequence number as $E_{Public \: key_{S}}(key_r+Sequence \: Number_S)$ by using the public key $Public \: key_{S}$ and the digital signature $E_{Public \: key_S}(E_{SK_R}(H(key_r+Sequence \: Number_S)))$. This way a MitM attacker can no longer fabricate the combination of session key $key_r$ and sequence number $Sequence \: Number_S$. $S$ can now decrypt the random string $key_r$ with the certificate sequence number $Sequence \: Number_S$ using $SK_{S}$ and also the digital signature by using $SK_S$ and $Public \: key_{R}$ respectively.
Now, $S$ and $R$ switch to the symmetric encryption. The recently established session key $key_r$ is used on both sides to encrypt and decrypt the message.



\section{Security Analysis and Correctness Sketch}\label{sect:correct}
In this Section we illustrate the arguments for the safety assurance implied by our protocol. The proposed protocol is resistant to MitM attack. The CA public key is conveyed to vehicles in secure settings. CA receives the request for the certificate deliverance and only the intended recipient will get the certificate $Cert$ from CA. An attempt to manipulate the certificate $Cert_S$ contents, in order to replace the attributes to fit the attacker vehicle attributes or the public key, will be detected as the digital signature $Sign_{CA}(H(Attribute_S+Public \: key_S))$ yields an impossibility to modify a certificate or to produce a totally new one. Receiver $R$ decrypts the digital signature using the CA pubic key $PK_{CA}$ and confirms the validity. Thus, any verifiable certificate has been originated by the CA and therefore the attributes coupled with a certain public key would uniquely characterize the vehicle.

\begin{table}
	\centering
            \begin{tabular}{ | p{1.2cm} | p{1.3cm} | p{1.8cm} | p{1.9cm} | p{3.3cm} | c | c | c | c | c |}
            \hline
               \textbf{Protocols} & \textbf{Direct iteration cost} & \textbf{Online authority interaction} & \textbf{Out-of-band verification} & \textbf{Coupling vehicle with communication} \\ \hline
               \textit{Proposed} & 2 rounds & No & Yes & Yes \\ \hline
               \textit{ISO-KE} & 3 rounds & Yes & No & No \\ \hline
               \textit{SIGMA} & 3 rounds & No & No & No  \\ \hline

            \end{tabular}
	\caption{Comparison with existing AKE protocols.}
	\label{tab:analyze}
	\end{table}

After the mutual authentication is done through a signed public key verification, coupled with the fixed sense-able attributes, a session key has to be established. A random string $key_r$ is generated at the receiver $R$ and is sent along with the certificate $Cert_R$, in response to sender's request for certificate $Cert_R$. As the $key_r$ can be replaced by a MitM, $S$ needs to authenticate the origin of $key_r$. Moreover, an attacker can manipulate the random string in between hence it requires an integrity verification mechanism. First, $R$ encrypts the $key_r$ and $Sequence \: Number_S$ using $S$ public key $Public \: key_{S}$, i.e. $E_{Public \: key_{S}}(key_r+Sequence \: Number_S)$ so that only $S$ can decrypt the random string using corresponding secret key $SK_{S}$. Thus, the confidentiality is ensured as only intended receiver can decrypt the $key_r$ as $D_{SK_S}[E_{Public \: key_{S}}(key_r+Sequence \: Number_S)]$. In order to verify the digital signature over $key_r$, a hashing algorithm $H$ is used to produce a hashed key string $H(key_r+Sequence \: Number_S)$. A digital signature $E_{Public \: key_S}(E_{SK_R}(H(key_r+Sequence \: Number_S)))$ is attached with $E_{Public \: key_{S}}(key_r+Sequence \: Number_S)$. Thus, integrity is maintained as only $R$ can generate these signature. Similarly, only $S$ can retrieve the $H(key_r+Sequence \: Number_S)$ from the signature using secret key $SK_S$ and $Public \: key_{R}$ as $D_{SK_S}(D_{Public \: key_R}(H(key_r+Sequence \: Number_S)))$. Next, the $H(key_r+Sequence \: Number_S)$ from digital signature is compared with the hashed key string generated locally. If both hashed key strings are similar, then the $key_r$ is accepted as a session key. Note that the signed, encrypted $key_r$ and $Sequence \: Number$ cannot be used as part of a replay attack, however, such usage will be detected by the sender and the receiver, as the actual value of $key_r$, is not revealed to the attacker. The use of synchronized date-time and signed association of the date-time can avoid even such unsuccessful attack attempts.

As per the Table~\ref{tab:analyze} the proposed approach is comparable to the existing AKE protocols such as ISO-KE~\cite{ISO/IEC-IS-9798-3} and SIGMA~\cite{DBLP:conf/crypto/Krawczyk03} mentioned above in Section~\ref{sect:AKE}. Accordingly, the first criteria of comparison is the iteration cost that determines the communication complexity. The proposed approach requires only two rounds of communication, i.e., sender-to-receiver and receiver-to-sender as part of authenticated key exchange. Furthermore, the next column indicate whether an online interaction is needed as online interaction is too restrictive and costly. Therefore, we design our scheme to eliminate the need of an active assistance from trusted third party for authentication. In contrast with other AKE protocols our proposed approach incorporates an out-of-band verification to cross-verify the vehicle authentication over wireless channel. In addition, by the use of out-of-band our scheme enables a coupling between the vehicle's certified identity and the authenticated key exchange over the wireless channel.





\medskip\noindent\textbf{Correctness.}
We next outline the widely known method of formal verification, i.e., Spi calculus which is an inherent derivative of Pi calculus. However, Spi calculus is adapted to security primitives and adversary model and provides an axiomatic proof of security~\cite{calculus}. The Spi calculus inherits certain powerful constructs from its ancestor Pi calculus and new cryptographic primitives have been added such as nonce, unique key, encryption, decryption, signing, and adversary process, etc. Secrecy, integrity and authentication are well motivated properties
for the application of Spi calculus. In addition, it is powerful in terms of testing equivalence, scope construct, assertion, predicates, adversary model and channel restriction that enables a synchronized communication among processes.

\smallskip
\noindent $\bullet$ \textit{Observational equivalence:} Accordingly, a formal ideal protocol description is normalized and combined with an adversary in terms of an independent process such that both formalizations of the ideal protocol leads to similar observational equivalences. It is useful for authentication as well as a secrecy property verification. For example, $A \approx B$ means that behaviour of the process $A$ and $B$ is indistinguishable and a third entity cannot identify the difference from running in parallel with any one of them. These testing equivalences are reflexive, transitive and symmetric.

\noindent $\bullet$ \textit{Trace analysis:} Apparently, trace-based reasoning is verified against valid communication sequences through message input and output. A protocol is considered secure if every trace resembles the ideal protocol phase, i.e. ideal sequence of communication. Spi calculus provides much deeper level of complexity and therefore freedom to achieve security goals in the presence of more complex adversary process.

\begin{table}[t]
  \centering
    \begin{tabular}{rl}
    $n$                & Name           \\
    $c$                & Communication channel        \\
    $u$, $v$, $w$, $x$, $y$, $z$, $t$, $l$    & Variable       \\
    $S$, $R$, $A$                     & Processes      \\
    $\overline{c}\langle N\rangle.$S       & Output process         \\
    $c$(\textit{x})$.$$S$           & Input process          \\
    $key^+$                     & Public key      \\
    $key^-$                      & Private key    \\
    $S$$|$$R$                       & Composition    \\
    (\textit{vn})$S$            & Restriction with bound $n$    \\
    $$[$M is N$]$S$                & Match          \\
    $S$$\simeq$$R$                           & Testing equivalence   \\
    \textit{Inst}$(M)$               & Instance of interaction     \\
    \end{tabular}
\caption{Notations used in Spi calculus}
\label{tab:label}
\end{table}

\smallskip
In what follows we would illustrate a formal realization of the proposed protocol using Spi calculus (see Table \ref{tab:label} for notations). Processes $S, R, A$ denote the communicating parties Sender, Receiver and Attacker, respectively. The rounds of message $M$ exchange between the sender and receiver called as one instance of the protocol and is denoted by \textit{Inst}$(M)$. Processes start exchanging the certificates in the following order.



First, in instance $Inst(Cert_S)$, sender $S$ sends the certificate $Cert_S$ to $R$, on channel $c_{SR}$ and $R$ receives the certificate on the same channel.

\scriptsize
\begin{align}
\BBB
S(Cert) \: &\triangleq \: \overline{c_{SR}}\langle Cert_S\rangle  \nonumber \\
S(Cert) \: &\triangleq \: \overline{c_{SR}}\langle Attribute_{S}+Public \: key_{S}||Sign[\{H(Attribute_{S}+Public \: key_{S})\}]_{CA^-}\rangle \nonumber \\
R \: &\triangleq \: c_{SR}(x).case \: x \: of \: y \nonumber \\
& \: \: \: \: \: \: let(y_1, y_2) \: = \: y \: in \nonumber \\
& \: \: \: \: \: \: case \: y_2 \: of \: [\{z\}]_{CA^+} \: in \: F(y_1) \nonumber\\
Inst(Cert) \: &\triangleq \: (vK_{CA})(S(Cert)|R) \nonumber
\BBB
\end{align}

\normalsize
Next, instance $Inst(Cert||key_r)$ executes in sequence, while receiver $R$ forwards certificate $Cert_R$ and session key $key_r$ to $S$.

\scriptsize
\begin{align}
R(Cert||key_r) \: &\triangleq \: \overline{c_{RS}}\langle {Cert_R}\rangle|\langle \{[key_r+Sequence \: Number_S]\}_{S^+} \nonumber \\
& \: \: \: \: \: \: ||\{[[\{H(key_r+Sequence \: Number_S)\}]_{R^-}]\}_{S^+}\rangle \nonumber \\
R(Cert||key_r) \: &\triangleq \: \overline{c_{RS}}\langle {Attribute_{R}+Public \: key_{R}||Sign[\{H(Attribute_{R}+{Public \: key}_R})\}]_{CA^-}\rangle \nonumber \\
& \: \: \: \: \: \: |\langle \{[key_r+Sequence \: Number_S]\}_{S^+}|| \nonumber \\
& \: \: \: \: \: \: \{[[\{H(key_r+Sequence \: Number_S)\}]_{R^-}]\}_{S^+}\rangle \nonumber \\
S \: &\triangleq \: c_{RS}(x)(u).case \: x \: of \: w \nonumber \\
& \: \: \: \: \: \: let(w_1, w_2) \: = \: w \: \nonumber \\
& \: \: \: \: \: \: in \: case \: w_2 \: of \: [\{v\}]_{CA^+} \: in \: F(w_1) \nonumber\\
& \: \: \: \: \: \: .case \: u \: of \: t \nonumber \\
& \: \: \: \: \: \: let(t_1, t_2) \: = \: t \: \nonumber \\
& \: \: \: \: \: \: in \: case \: t_2 \: of \: [\{\{[l]\}_{S^-}\}]_{R^+} \: in \: F(t_1) \nonumber \\
Inst(Cert||key_r) \: &\triangleq \: (vK_{S})(vK_{R})(R(Cert||(key_r+Sequence \: Number_S))|S) \nonumber
\end{align}

\normalsize
We analyze the authenticity and secrecy properties in Claim 4.1 and 4.2, respectively.

~\\
\noindent\textbf {Claim 4.1} \textit{The proposed session key establishment protocol respects the authenticity property i.e. $F(y_S)$ a local function computation at $R$, is accepted, if indeed it arrived from $S$.}

\noindent\textit{Proof.} According to the property of \textit{authenticity}, the receiver is able to verify that the certificate is indeed, from the sender, that the certificate claims to come from. Here, we prove the authenticity of the certificate $Cert_S$ and the sender $S$, before the local function computation $F(Cert_S)$ at $R$.

First instance of the certificate exchange, while certificate moves from sender to receiver is as follows:

\scriptsize
\begin{align}
Inst(Cert) \: &\triangleq \: (vK_{CA})(S(Cert)|R) \nonumber
\end{align}

\normalsize

In order to satisfy, the property of authenticity, following statements hold true for the first instance:

\begin{itemize}
    \item The recipient can verify, that the certificate $Cert_S$, indeed originated at the CA. Because the receiver holds the CA public key i.e. $CA^+$, and is able to verify the CA signature over the hashed certificate contents, provided that the condition below holds true: $$[(case \: y_2 \: of \: [\{z\}]_{CA^+}) \: = \: y_1]$$
        Thus, after the step 2(a) in Figure \ref{fig:dcv}, receiver $R$ knows that the \textit{digitally signed} certificate $Cert_S$ holds valid contents regarding the sender $S$. Subsequently, the receiver authenticates the certificate $Cert_S$.
    \item The certificate $Cert_S$, is attributed to the actual sender $S$, if and only if $F(y_1)$ qualifies the out-of-band verification. Because, the receiver extracts the \textit{authenticated certified attributes} of the sender vehicle $S$. Then it verifies the fixed out-of-band channel attributes and confirms, that the authenticated attributes still hold true.
    \item The receiver $R$, derives the \textit{coupled public key} from $Cert_S$, and knows that it indeed belongs to the certified attribute holder, if the condition below is satisfied: $$[(case \: S^+(y_2) \: of \: [\{z\}]_{CA^+}) \: = \: (S^+(y_1))]$$
\end{itemize}

\normalsize
For the second instance, receiver replies back with the certificate $Cert_R$ concatenated with the hashed and signed session key $(key_r+Sequence \: Number_S)$.

\scriptsize
\begin{align}
Inst(Cert||key_r) \:  &\triangleq \: (vK_{S})(vK_{R})(R(Cert||(key_r+Sequence \: Number_S))|S) \nonumber
\end{align}

\normalsize
Second instance holds true on the following properties, while analyzing the property of authenticity.

\begin{itemize}
  \item The sender $S$ can verify, that the certificate $Cert_R$, indeed originated at the CA, if the condition below holds true:
  $$[(case \: w_2 \: of \: [\{v\}]_{CA^+}) \: = \: w_1]$$
  Because the sender $S$ holds the CA public key, i.e., $CA^+$, and is able to verify the CA signature over the hashed certificate contents. Thus, after the step 3(a) in Figure \ref{fig:dcv}, sender $S$ knows that the \textit{digitally signed} certificate $Cert_R$ holds valid contents regarding the receiver $R$. Consequently, the sender authenticates the certificate $Cert_R$.
  \item The certificate $Cert_R$, is attributed to $R$, if and only if, $F(w_1)$ qualifies the out-of-band verification. Because the sender extracts the \textit{authenticated certified attributes} of the receiver vehicle $R$. It then verifies the fixed out-of-band channel attributes and confirms that the authenticated attributes still hold true.
  \item The sender $S$, derives the \textit{coupled public key} from $Cert_R$, and knows that it indeed belongs to the certified attribute holder $R$, if the condition below holds true: $$[(case \: R^+(w_2) \: of \: [\{v\}]_{CA^+}) \: = \: (R^+(w_1))]$$
  \item The \textit{binding} between the session key $key_r$ and the certificate $Cert_R$ holds true. It requires that $[(case \: t_2 \: of [\{\{[l]\}_{S^-}\}]_{R^+}) \: = \: t_1]$, provided that the condition $[(case \: w_2 \: of \: [\{v\}]_{CA^+}) \: = \: w_1]$ is also verified. Because, it is confirmed that the signature over hashed session key, utilizes the secret key $R^-$, and can only be generated by $R$.
\end{itemize}

The following claim proves the second property, \textit{secrecy perseverance} of the proposed protocol.
\begin{flushright}
  $\blacksquare$
\end{flushright}

~\\
\noindent\textbf {Claim 4.2} \textit{The proposed session key establishment protocol respects the secrecy property. Any instance of certificate exchange does not reveal the secret session key and subsequently, any instance of the session key encrypted message exchange does not reveal the message contents.}

\noindent\textit {Proof.} According to the \textit{secrecy} property, an attacker cannot distinguish the \textit{different} messages encrypted with the same or different session key, for the same or different pair of vehicles/processes. The message must be revealed to the intended recipient only.
First we prove the secrecy property for the session key exchange between the sender and receiver, and then for the message exchange within the shared key session.

In the first instance, the sender exchanges the certificate; we do not assume a secret certificate exchange. In the second instance, we need the secrecy regarding the session key exchange, in order to ensure the secrecy of session key encrypted messages in the current session.

\begin{itemize}
    \item The session key exchange in the second instance is a secret, between the receiver $R$ and sender $S$, as the $[(case \: t_2 \: of \: [\{l\}]_{S^-}) \: \simeq \: (case \: t'_2 \: of \: [\{l'\}]_{S^-})]$. Only the sender knows the secret key $S^-$ and is able to verify the signature.
    \item The hashed and signed sequence number $Sequence \: Number_S$, ensures that it was indeed sent to $S$ and the condition below must hold true: $$[(case \: t_2 \: of \: [\{\{[l]\}_{S^-}\}]_{R^+}) \: = \: t_1]$$
        Only, the secret key $S^-$ holder, i.e., $S$ can decrypt and verify the signature over the session key and sequence number, see step 3(b) in Figure \ref{fig:dcv}.
\end{itemize}

After the secure session key exchange as stated above, current session messages are encrypted with the secret session key.
The current session key is a shared secret between the sender $S$ and receiver $R$, only. It is also important to mention that the local function computation $F(t_1)$ at $S$ is inherently secure, and does not reveal the deciphered message contents local to $S$. Hence, we can say that the session key is securely exchanged with $S$ followed by the secure $F(t_1)$ computation at $S$. Thus, future messages of the current session, are secretly shared between the two, as they are encrypted using a unique shared secret key.
\begin{flushright}
  $\blacksquare$
\end{flushright}

\section{Conclusion and Future Work}\label{sect:conclusion}
The proposed work provides Man-in-the-Middle (MitM) attack resistance and mutual authentication using certified public key and out-of-band sense-able attributes. As the Certificate Authority (CA) preprocesses every vehicle's public key and the unchangeable visual attributes, there is no way that MitM can fake the public key or the unchangeable attributes. Also, the out-of-band attributes are sense-able and can be confirmed while moving on the road. There is no need to communicate with the CA during the real-time session key establishment of a secret key for the mutual authentication of vehicles. The proposed protocol is simple, efficient and ready to be employed in current and future vehicular networks. More sophisticated scheme that specifically requires additional communication hardware, which is not currently available in vehicles, may also verify dynamic attributes in case the adversary is able to clone the vehicle with license number~\cite{ncadyna}.


\noindent\textbf{Acknowledgment.}
We thank Niv Gilboa, C. Pandu Rangan, Sree Vivek, anonymous reviewers and the editor for valuable comments.


\end{document}